# A DERIVATION OF BELL'S INEQUALITIES IN WIGNER FORM FROM THE FUNDAMENTAL ASSUMPTION OF STATISTICAL MECHANICS

*Ian T. Durham[†]*
[†]*Department of Physics*
*Saint Anselm College*
*100 Saint Anselm Drive, Box 1759*
*Manchester, NH 03102*
*idurham@anselm.edu*

**Abstract**

Bell's inequalities in the form given by Wigner are derived from the so-called fundamental assumption of statistical mechanics. I then demonstrate the possible relationship between these inequalities and the second law, particularly if assuming the Pauli exclusion principle dictates the expected outcomes.

## I. BACKGROUND

Bell's inequalities [1] form the basis of a considerable amount of work in modern quantum mechanics since they give a mathematical view of the phenomenon of entanglement. In particular, these inequalities form one of the conceptual foundations of quantum information and computation [2] and include versions formed using entropies [3]. A very straightforward and simple derivation of a general form of Bell's inequalities was given by Wigner [4], described in detail in Sakurai [5], whereby two independent observers make a series of spin alignment measurements on particles in what essentially amounts to independent sequential Stern-Gerlach devices fed by a single reservoir. An inequality leading to the final Bell inequality is derived from the basic positive, semi-definite nature of the populations of the pairs of particles that are involved. Locality is implicitly in this derivation as is the very quantum assumption arising from the Pauli exclusion principle that correlated pairs cannot have the same spin alignment (since they cannot share the same set of quantum numbers).

Numerous experiments over the years have realized violations of Bell inequalities beginning with Aspect [6] and continuing to now. For example, an alternate derivation of these equations (there are several), predicts a very specific value for this violation that has been repeatedly verified in laboratory experiments [7]. Debate as to the precise meaning of these violations is ongoing with some arguing that quantum mechanics is inherently nonlocal [8].

There is a subset of research, mostly theoretical in nature, focused on linking entanglement to thermodynamics in some way [9]. In particular, Horodecki, Oppenheim, and Horodecki [10] have argued that the laws governing entanglement may well be thermodynamic in nature, or, at the very least, possess thermodynamic corollaries.

In this paper, I derive Bell's inequalities in the form found by Wigner from basic arguments about statistical mechanics including the so-called fundamental assumption of statistical mechanics. In order to do this I must first give the definitions from which I will be working, most of which should be familiar.

## II. STATISTICAL ASSUMPTIONS AND INTERACTING SYSTEMS

In any given system consisting of an aggregate of members (e.g. an ideal gas) there are often any number of ways in which the members can arrange themselves to form a given macroscopic state, called a macrostate. Each way in which the system can orient itself (each configuration) is called a microstate and the number of such microstates for a given macrostate is called the multiplicity of the system for that macrostate. So, for instance, given a pair of dice – one red, one black – say we are interested rolling a seven. There are, of course, numerous ways we can get a seven – one die could be a six with the other a one, or one die could be a three while the other is a four. In addition, the red die could be a six, for instance, while the black die is a one, *or* the black die could be a six while the *red* die is a one. There are many ways to describe this using the given terminology, but we would call the roll of seven the macrostate of the combined system (or sometimes the macropartition though this is technically related to interacting systems). The multiplicity is then simply the number of possible microstates for that given macrostate (in this case six). The probability of rolling a seven, say, is simply the ratio of the multiplicity of this macrostate to the total multiplicity. For a pair of dice, the total multiplicity is 36 and thus the probability of rolling a seven is one-sixth. Conversely, there is only *one* way to roll a two and thus a probability of one-thirty-sixth.

For thermodynamic systems supplied by an infinite reservoir, it is assumed that all *accessible* macrostates are equally probable. This is known as the fundamental assumption of statistical mechanics. Note that there is no analogy to rolling a pair of dice since, in the long run, all the macrostates for a pair of dice are *not* equally probable. It does work, however, for coin tossing.

In any case, it is instructive to define the entropy of a system in terms of the multiplicity as

$$S \equiv k_B \ln \Omega \qquad (1)$$

where $k_B$ is Boltzmann's constant.[*] Ultimately this is simply a more convenient way to represent multiplicities since, for very large systems, $\Omega$ can become unwieldy. The natural log simply makes the number more manageable and Boltzmann's constant is simply a conversion factor. In fact, in this sense it really has nothing to do with disorder [11] unless disorder is taken to mean how many configurations the system can have. Equation (1) also implies that multiplicities can be expressed in terms of entropy by $\Omega = e^{S/k}$.

---

[*] Note that many authors define entropy to be $S \equiv -k \sum_s P(s) \ln P(s)$ where $P(s)$ is the probability that the system will be in microstate *s*.

Assume, then, two interacting systems *A* and *B* with multiplicities $\Omega_A$ and $\Omega_B$. The multiplicity of the combined system, since it is simply the total number of configurations the system can have, is

$$\Omega_{AB} = \Omega_A \Omega_B = e^{S_A/k} e^{S_B/k} = e^{(S_A + S_B)/k}. \tag{2}$$

Note that this implies the additivity of entropy and follows from a purely statistical argument (e.g. there are six possible outcomes on the roll of a single die but thirty-six possible outcomes on the roll of a *pair* of dice).

## III. SEQUENTIAL MEASUREMENTS

Imagine, then, a source for spin measurements, for example an oven feeding a Stern-Gerlach device or series of devices [12], that emits particles in such a way that two observers, Alice and Bob, perform independent measurements along three, not necessarily orthogonal axes, $\hat{\mathbf{a}}$, $\hat{\mathbf{b}}$, and $\hat{\mathbf{c}}$. The same assumptions are made here that are made in [12] but note that this implies that the exclusion principle dictates what we expect the measurement results to be. So, for instance, if Alice and Bob are always measuring pairs of correlated particles, if Alice measures $+\hat{\mathbf{a}}$ for particle 1, Bob *must* measure $-\hat{\mathbf{a}}$ for particle 2. Since this is inherent in the derivation in [12] and it is a quantum property, Bell's inequalities as derived in this manner are not entirely classical. It is an important distinction that I will discuss further in a moment.

Given, then, a series of measurements by Alice and Bob, all particle pairs fall into one of eight populations summarized in Table 1. For a single particle pair this would make only eight possible configurations which I will call the macrostates (similar to one of the *microstates* for the example of the dice – the difference will become clear in a moment). Let's say that our source is not entirely random and is somehow biased towards one of these macrostates. Perhaps there are *N* particles in this particular macrostate but much, much fewer particles in the other macrostates. This macrostate must be more likely to occur in a subsequent measurement than the other macrostates which suggests it has a higher multiplicity. Let's assign a multiplicity to each of the populations described in Table 1, then.

So, say we are interested in the specific state in which Alice measures $+\hat{\mathbf{a}}$ and Bob measures $+\hat{\mathbf{b}}$. From Table 1 we see that this corresponds to populations $N_3$ and $N_4$ and we can assign each a multiplicity, $\Omega_3$ and $\Omega_4$ respectively. The joint multiplicity for this combined state is

$$\Omega_{34} = \Omega_3 \Omega_4. \tag{3}$$

Now, the fundamental assumption of statistical mechanics states that all accessible macrostates are equally probable in the long run. In our example here, this means we have two boundary conditions that must apply: first the reservoir (e.g. the oven) must be infinite and the ensemble within the reservoir must be entirely random. If this holds, in the long run all the multiplicities for all the possible populations are roughly equal since we're assuming these numbers are at least partially dependent on the number of particles in a given population (I will have more to say on this assumption a bit later). So, in essence we are assuming that $\Omega_1 \approx \Omega_2 \approx \Omega_3 \approx ... \Omega_n$. As such we can write inequalities of the form

$$\Omega_{34} \leq \Omega_{24} + \Omega_{37}. \tag{4}$$

Note, however, that this inequality only holds in the limit of a completely randomly oriented infinite reservoir since it requires that all the individual multiplicities are roughly equal.

The probability, then, that Alice will measure $+\hat{\mathbf{a}}$ and Bob will measure $+\hat{\mathbf{b}}$ is

$$P(+\hat{\mathbf{a}};+\hat{\mathbf{b}}) = \frac{\Omega_{34}}{\sum \Omega_n} = \frac{\Omega_{34}}{\Omega_{total}} \tag{5}$$

where the denominator represents the sum total of all possible configurations (total multiplicity). It is straightforward then to rewrite equation (4) as

$$P(+\hat{\mathbf{a}};+\hat{\mathbf{b}}) \leq P(+\hat{\mathbf{a}};+\hat{\mathbf{c}}) + P(+\hat{\mathbf{c}};+\hat{\mathbf{b}}) \tag{6}$$

which is the form Wigner found for Bell's inequalities ([5] and [12]).

## IV. THE SECOND LAW AND THE PAULI PRINCIPLE

What if the reservoir is *not* infinite? Imagine, for instance, a bag of marbles – blue, black, white, and red – initially many of each color randomly distributed within the bag. At the start, if the populations of the different colored marbles are roughly equal, the probability of pulling out, say, a red marble, is equal to the probability of pulling out, say, a black marble. But say after a good run of picking we know we've seriously depleted the number of red marbles in the bag. Subsequently, the probability that we will extract a red marble might be substantially *less* than the probability of finding a black marble. Another way of saying this is to say that as we continue to pick marbles, the probabilities of picking a certain color converges to the point where, just before we pick the final marble, it is 100% likely that we will pick the only color that is left. This, in fact, is analogous to one form of the second law of thermodynamics in which the probabilities for various microstates increase for unrealized microstates as a system passes through these states enroute to equilibrium [13]. In fact this argument for the second law is merely a strong argument about the behavior of probabilities and is the reason the second law is not, in fact, fundamental [14]. This is precisely why our assumption of an infinite reservoir was so important above. If the reservoir were not infinite it is conceivable that the actual probabilities for a given configuration might be different and Bell's inequalities, as represented in equation (6), could not be formed. Also note that our derivation, then, *assumes* that the configurations of the individual correlated pairs is somehow *predetermined*. This implies that violations of these inequalities is somehow also a violation of determinism which we interpret as locality. Of course, quantum mechanics via most interpretations assumes just the opposite: that they are not determined until they are actually measured.

Another point that I wish to make, that I briefly mentioned before, is that the Pauli exclusion principle is inherent in both my derivation and Wigner's original since it is assumed that if Alice measures $+\hat{\mathbf{a}}$ on particle 1, Bob *must* measure $-\hat{\mathbf{a}}$ on particle 2 or find complete randomness in his results [5]. But this is a purely *quantum* assumption. In a sense, since correlation is assumed, violations of the inequality deal with *locality* as opposed to correlation in general. As such, (quantum) correlation and entanglement are two slightly different things, with the latter combining correlation itself with seemingly non-local behavior.

From a classical standpoint, then, equation (4) isn't quite *physically* right. If we remove the quantum restriction that Bob can't measure $+\hat{\mathbf{a}}$ on particle 2 if Alice measures $+\hat{\mathbf{a}}$ on particle 1, and assume that the right-hand side of equation (4) can be interpreted as an and/or situation (since the individual multiplicities are assumed to be equal), the multiplicity for Alice measuring $+\hat{\mathbf{a}}$ and Bob measuring $+\hat{\mathbf{c}}$ *or* Alice measuring $+\hat{\mathbf{c}}$ and Bob measuring $+\hat{\mathbf{b}}$, is $\Omega_{2437} = \Omega_2\Omega_4\Omega_3\Omega_7$. Equation (4) could then be more properly expressed as

$$\Omega_{34} \leq \Omega_{2437}. \tag{7}$$

Taking the natural log of both sides of equation (7) produces an inequality formed out of the additivity condition for entropy of the form

$$S_3 + S_4 \leq S_2 + S_4 + S_3 + S_7. \tag{8}$$

As I previously mentioned, inequalities have been also formed for Shannon entropies [3]. In a sense it could be argued that equation (8) is simply another way of stating the second law of thermodynamics since the second law includes the implication that, at least classically, negative entropies do not exist[*] (which follows from the purely statistical definition of entropy). Note, however, that it is not entirely clear that equation (8) can lead directly to equation (6). If the entropies are simply another way to keep track of the multiplicities, ratios of entropies would be equally useful in forming probabilities in which case equation (8) would lead directly to equation (6). If this were the case, violations of Bell's inequalities would seemingly violate the second law, something that would have profound implications for the realization of practical quantum computers. Note, however, that this depends greatly on the actual interpretation of entropy.

Finally, note that I have proposed an explanation for this seeming incongruity [15]. Essentially I point out that Bell's inequalities are based on classical assumptions that cannot be applied to quantum systems, most notably because many quantum systems have off-diagonal and/or negative terms in their density matrices and these are the most likely cause of violations of inequalities of this form.

---

[*] Negative entropies have been shown to exist in quantum mechanics, e.g. see [3].

| Population | Particle 1 | Particle 2 |
|---|---|---|
| $N_1$ | $(+\hat{\mathbf{a}},+\hat{\mathbf{b}},+\hat{\mathbf{c}})$ | $(-\hat{\mathbf{a}},-\hat{\mathbf{b}},-\hat{\mathbf{c}})$ |
| $N_2$ | $(+\hat{\mathbf{a}},+\hat{\mathbf{b}},-\hat{\mathbf{c}})$ | $(-\hat{\mathbf{a}},-\hat{\mathbf{b}},+\hat{\mathbf{c}})$ |
| $N_3$ | $(+\hat{\mathbf{a}},-\hat{\mathbf{b}},+\hat{\mathbf{c}})$ | $(-\hat{\mathbf{a}},+\hat{\mathbf{b}},-\hat{\mathbf{c}})$ |
| $N_4$ | $(+\hat{\mathbf{a}},-\hat{\mathbf{b}},-\hat{\mathbf{c}})$ | $(-\hat{\mathbf{a}},+\hat{\mathbf{b}},+\hat{\mathbf{c}})$ |
| $N_5$ | $(-\hat{\mathbf{a}},+\hat{\mathbf{b}},+\hat{\mathbf{c}})$ | $(+\hat{\mathbf{a}},-\hat{\mathbf{b}},-\hat{\mathbf{c}})$ |
| $N_6$ | $(-\hat{\mathbf{a}},+\hat{\mathbf{b}},-\hat{\mathbf{c}})$ | $(+\hat{\mathbf{a}},-\hat{\mathbf{b}},+\hat{\mathbf{c}})$ |
| $N_7$ | $(-\hat{\mathbf{a}},-\hat{\mathbf{b}},+\hat{\mathbf{c}})$ | $(+\hat{\mathbf{a}},+\hat{\mathbf{b}},-\hat{\mathbf{c}})$ |
| $N_8$ | $(-\hat{\mathbf{a}},-\hat{\mathbf{b}},-\hat{\mathbf{c}})$ | $(+\hat{\mathbf{a}},+\hat{\mathbf{b}},+\hat{\mathbf{c}})$ |

**Table 1**. Spin-component matching for Alice and Bob, adapted from ref. 5, p. 229.